# Astronomy and landscape at the prehistoric settlement

# *Villaggio dei Faraglioni*, Ustica, Sicily

**Franco Foresta Martin[1], Giulio Magli[2]**

[1]*Laboratorio Museo di Scienze della Terra Isola di Ustica*
[2]*Faculty of Civil Architecture, Politecnico di Milano*

**ABSTRACT**

The placement of the Middle Bronze Age settlement Villaggio dei Faraglioni on the Ustica island, one of the best preserved prehistoric village of the Mediterranean area, is analyzed from a cognitive point of view, taking into account archaeoastronomy and landscape archaeology aspects. It turns out that the place might have been selected because of its privileged position with respect to the landscape, better than conforming to utilitarian/defensive considerations, as instead occurs for contemporary, for instance Mycenaean, sites. From the village it was indeed possible to follow the rising and setting of the sun in the months close to the winter solstice on the two elevated peaks existing on the opposite side of the island, in a symmetric way. As a consequence, the inhabitants could determine the day of the winter solstice with an approximation of a few days, using this "partial calendar" for agricultural and navigational purposes. A possible astronomical orientation of the urban layout is also analyzed.

## INTRODUCTION

In recent years, Archaeoastronomy has developed from the mere analysis of possible astronomical alignments at a site to a quite sophisticated branch of cognitive archeology, which tries to understand the relationship of the ancient landscapes - and of the people who conceived them - with the sky and with the celestial cycles (see e.g. Magli 2015 and references therein). Adopting this broad viewpoint, we study here one almost untouched, very peculiar landscape of the Middle Bronze Age: that of the tiny island of Ustica, in the Mediterranean sea. It turns out that the prehistoric settlement of the island, Villaggio dei Faraglioni, is located in the unique possible position from where it was (and is) possible to use the terrestrial and uneven horizon as a means of controlling the cycle of the sun during winter. Since the position of the settlement is quite uncomfortable and, in turn, it does not fully meet utilitarian, or defensive, criteria, we propose that it was the relationship with the landscape and the sky the main reason for the choice of the settlement's place.

## THE FARAGLIONI PREHISTORIC VILLAGE

The island of Ustica (Fig. 1) is located in the Mediterranean Sea, 70 Kms to the north of Palermo. It is a tiny land of volcanic origin stretching about 2.7 km. by 4.5 km.

Its geological history is quite complex, as it is the result of different volcanic eruptions, effusive and explosive, and of their consequences. The remnants of three main sub aerial volcanoes form today a rib hill which traverses the island in a roughly east-west direction: Monte Guardia dei Turchi, the first emerged volcano, is the tallest (248 m asl) and occupy a central position; Monte del Fallo (234 asl) to the west; Monte Falconiera (157 m asl) to the East. To the north of this ridge, a flat, fertile land – Piano di Tramontana – was formed by the transgression of the sea on the preexisting lavic material, while to the south the hills open into the unique port of the island, Cala S. Maria, and into others small marine terraces that characterize the southwest sector too.

The island was populated already since the 6th millennium BC, although not continuously. During the Middle Bronze Age, between 3400-3200 yrs BP, it was intensely and permanently inhabited (Mannino 1979). There exist traces of small Middle Bronze Age dwellings in the eastern zone of the island at Punta dell'Omo Morto (at the feet of Falconiera hill) and at Case Vecchie (upstream of the Ustica town); in the western area of the island at Spalmatore (near the namesake touristic village); in the southern area of the island at Piano dei Cardoni and at San Paolo. Others small settlements were located on some peaks of the hill's rib, like the Culunnella Village located on the eastern flank of Monte Guardia dei Turchi (Spatafora and Mannino 2008). But the unique conspicuous settlement of this period is that located directly to the sea on the rock terrace of Piano di Tramontana and today called Villaggio dei Faraglioni ("faraglione" means a rock spur which emerges from the sea nearby).

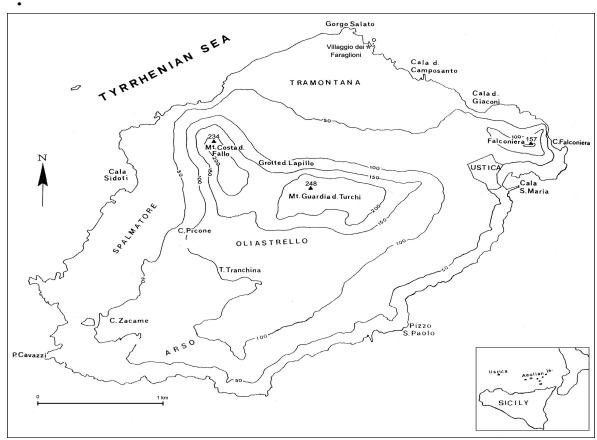

**Figure 1. Map of Ustica, showing the position of the prehistoric settlement.**

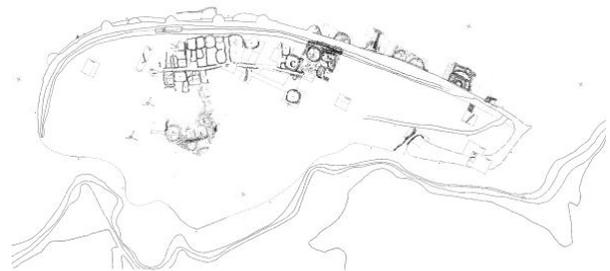

**Figure 2. Map of the fortification wall and of the excavated parts of the prehistoric village (courtesy F. Spatafora).**

Several different archaeological excavations, carried out since Seventies, although discontinously, by G. Mannino (1970, 1979, 1982), R. Ross Holloway & S. Lukesh (1995, 2001) and F. Spatafora (2004, 2008) highlighted a settlement that has been defined one of the best-preserved Middle Bronze Age town of the Mediterranean region (Counts and Tack 2009). Since only a part of these works are in English and easily accessible, we give here a brief account of the main points which have been uncovered by excavations.

The village is enclosed in a massive fortification wall (Fig. 2) which delimits an area of about 7000 square meters naturally protected by the sea on the opposite sides; there, the rocky cliff has an height of about 20 m asl. Inside the wall a certain number of dwellings organized along main longitudinal walk roads have been unearthed (Fig. 3,4).

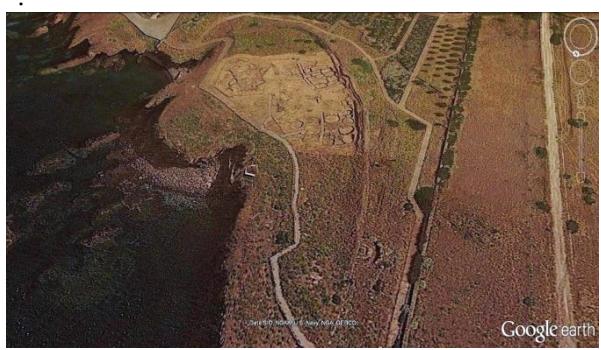

**Figure 3. Satellite view of the prehistoric village from the north (courtesy Google Earth Pro)**

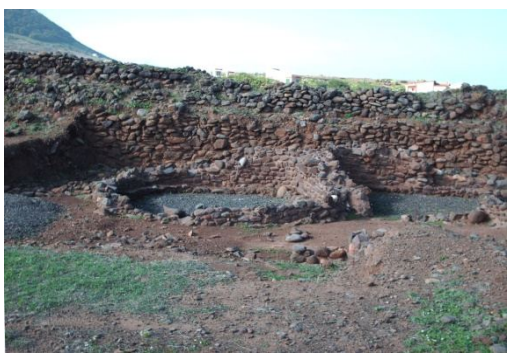

**Figure 4. One of the circular dwellings, with the fortification wall immediately behind.**

A tiny cove is located some hundreds meters to the west and may have served as a harbour for small boats. According to some authors the inhabited, wall-circled area was greater than today and a landslide occurred on the seaside, insulating the biggest "faraglione" named Colombaio that was once connected to the mainland. The distribution of huts, courtyard and streets suggest a well ordered organization of the space inside the village, a sort of "urbanistic plan" similiar to that found in the coeval village of Thapsos, in the Magnisi peninsula, near Syracuse, Eastern Sicily. The huts of the Faraglioni Village were circular, oval, or rectangular with rounded corners and were placed next to one another, at the sides of streets, which are about one meter wide. Between groups of adjacent huts, common courtyards were made, in order to place containers for food and big pitchers for the collection of rainwater, the latter being of vital importance in an island completely devoided of water spring. The furnishing unearthed inside the huts was rich and well preserved. Some vases recalls the style of coeval ceramics found in the Milazzese Village of Panarea, in the Eolian Island. Some other vases are in typical Thapsos Style, consisting in deep truncated cone bowls on high trumpet feet, probably used to consume meals sitting on the ground. A lot of cups, bowls, pitchers, cruses and little amphoras were extracted from the floor or from the collapsed part of the huts. A big hut placed in the current central part of the village exhibit an internal anular bench and has been interpreted as an area devoted to worship. In the southern section, inside a well limited space, five stone moulds for the fusion of metal tools were recovered, suggesting the existence of a handicraft workshop.

The site had several distinct phases of occupation, starting at about 1400 BC and ending with the late Bronze Age, around 1250/1200 BC. It was inhabited by a community devoted to fishing, agriculture and sheep farming, but also well inserted into the Mediterranean trade, as several foreign contacts are attested to. In particular, the connection with the contemporary Sicilian culture of Thapsos is self-evident in ceramics (Voza 1972); worked obsidian – a material which is not native to the island – was imported from Lipari and Pantelleria (Tykot 1995, Foresta Martin 2013). Evidence of long-distance contact was also found. A single fragment of Mycenean ceramic and a few necklace beads in glass paste attest relationship with populations of the continental Greek Bronze Age. Moreover, some sherds with incised decoration in Appennine style document the partecipation of the Island in the Tyrrhenian traffic with the italian peninsula. Putting all together, the rational urbanistic plan of the village and the wealth of furnishing of the huts testify a well-structured social and economic organization, as well as a high standard of living of the inhabitants. Vast sectors of the village are still unexplored however, and much remains to be unearthed about the life of the inhabitants, as well as about their funerary practices and their suddenly disappearing around 1200 BC. The Village was indeed abruptebly left, leaving all the belongings in homes. Two hypotheses are advanced for this sudden flight: a hostile invasion or a natural disaster that induced the population to find a safer place. In any case, the site where the village was founded does not look as the optimal place for a permanent settlement for a long list of reasons. First of all, it is located on the northern lowland, while it would have been relatively easy to fortify, with a comparable building effort, an elevated position on the hills. Second, the view to the whole arc of the southern horizon (and thus towards Sicily) is completely obstructed by the previously mentioned hills, so communication (by fire signals, or other means) with elevated outposts was mandatory if – as it is likely of course – the inhabitants wanted to control the maritime traffic in front of the island (Ustica is visible from Palermo in clear days and viceversa). Not last, the settlement is located just above the sea; the rock terrace of the coast is not high enough to protect the area from the violent winds and sea storms which are typical on the island, especially during winter.

Another interesting observation is the following. There are conspicuous examples of bronze Age and Iron Age fortifications which made intelligent use of promontories or high terraces to integrate natural defenses with the addition of walls. For instance, this occurs in a settlement which is fully contemporary to Ustica, the late-Mycenaean (1250-1150 BC) site of Maa (Palaiokastro) in Cyprus. Maa is in fact a small peninsula, naturally protected by the sea, whose land side was fortified with a transverse stone wall. Again, the builders at Ustica – although they did use the sea as a defensive boundary - did not apply this kind of utilitarian rule, since the area protected by the stone wall develops in a direction roughly parallel to the coast, and as a consequence the fortification is much longer than it would have been, for instance, if the village would have been placed at the northernmost offshoot of the island, which actually is just a few hundred of meters further to the north-west.

The above observations are certainly enough to justify a new look at the settlement – whose ancient landscape can fortunately be considered as intact – from a cognitive point of view and, in particular, from the point of view of the relationship with the sky.

## ASTRONOMY AND LANDSCAPE AT FARAGLIONI

To investigate the relationship of the prehistoric village with the peculiar landscape it is merged in and with the sky, we have measured the whole site with a combined compass-clinometer instrument, aquiring thus standard data of use in archaeoastronomy: azimuth counted positively from true north and horizon height, for each relevant direction. Corrections for magnetic declination have of course been applied, reaching an estimated accuracy of ½°. The data have been controlled using Google Earth Pro. The corresponding declinations have been calculated using the program Get-DEC kindly provided by Clive Ruggles, which takes into account refraction. The results, reported in Table 1, show that:

1) The urban layout is characterized by a main street which runs parallel to the fortification wall. The street and the wall exhibit a bend to the north, after the bend the street has a bearing ~162° with horizon height 3°, yielding a declination -44° 30'. This declination corresponds, with an error less than 20', to the rising of the bright star Rigil Kent in the 13th century BC. The circumstance is impressive, because interest for the stars of the Crux-Centaurus group is very well attested in contemporary Mediterranean cultures such as Nuragic Sardinia (Belmonte and Zedda 2004) and Talayotic Menorca (Hoskin 2001). We stress, however, that the wall and the village layout are (roughly) parallel to the coastline and built on an flat area so it is difficult to decide which direction (wall or urban axis) was planned first and why.

2) The location chosen for the village is very peculiar. This is especially evident from a point which is also the most obvious observation point to be taken namely, the center of the circular dwelling (Room 12) which has been proposed by excavators as the likely sacred and political center of the village. It happens that from this point (denoted by C in Fig. 2) the easternmost Monte Falconiera and westernmost Monte Costa del Fallo hills coincide with the extreme points visible of the island at the horizon, with Monte Guardia dei Turchi hill in the middle; the remaining horizon being occupied by the sea. Further, the two peaks are seen in a symmetrical fashion with respect to due south. The declinations corresponding to the two peaks are indeed -23°12' and -22°58' respectively. The declination of the sun at the winter sosltice in 1300 bC (due to the slight variation of the obliquity of the ecliptic) was slighlty less than today, precisely of -23°50', so the difference is of 0°38' and of 0°52' respectively. Taking into account the solar diameter (30') this means that it was (and actually is still today) possible to see the midwinter sun rising on the peak of Monte Falconiera and setting on the peak of Monte Costa del Fallo, so that the arc of the Sun in the shortest day of the year embraces the whole heartland of the island, as seen from the Village; a fact that suggests a symbolic value for the placement of the Village itself (Fig. 5). The spectacle is quite impressive, as we verified on site at December solstice 2015 (Fig. 6,7). These pictures can be considered as quite close representations of what could be seen 3200 years ago or so

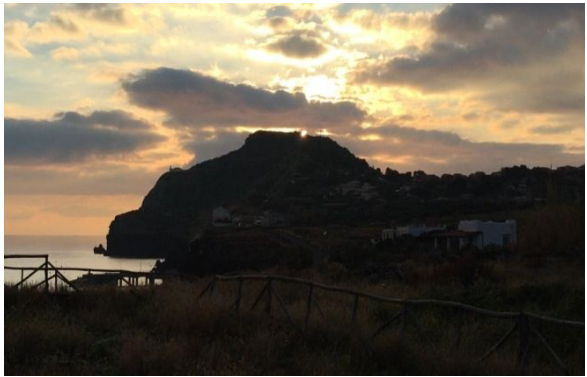

**Figure 5. Ustica, Villaggio dei Faraglioni, winter solstice 2015. The sun rises from behind Monte Falconiera**

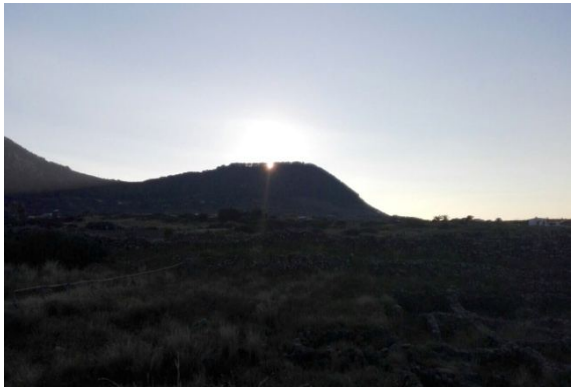

**Figure 7. Ustica, Villaggio dei Faraglioni, winter solstice 2015. The sun sets behind Monte Costa del Fallo.**

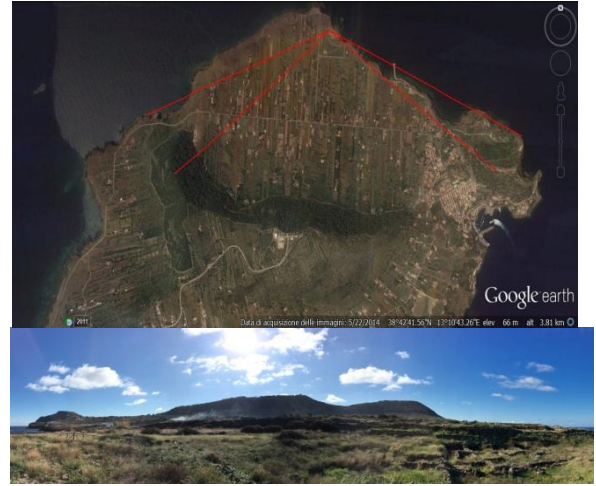

**Figure 8. Ustica, Villaggio dei Faraglioni. Panoramic view of the southern horizon as seen from the Prehistoric Village.**

**Figure 6. Google Earth Pro image, showing the rising/setting azimuths of the sun at winter solstice and the azimuths of the first/last rising and setting of the sun on the Island, as seen from the village. The corresponding data are given in Table 1. (Image courtesy Google Earth, drawings by the authors)**

Further, the extreme points of the two hills repectively to the south-east and the south-west yield declinations -19° and -16° respectively. The calculated dates (Gregorian) in which it was possible to see the sun "reach" and "leave" the island to the east are November 18 and January 24. Curiously, around the same date in November the star Rigil had heliacal rising in the 13th century BC. The calculated days to the west are November 6 and February 5.

## DISCUSSION

To summarize, we can say that the choice made for the placement of the Middle Bronze Age settlement Villaggio dei Faraglioni on the Ustica island is quite puzzling, both from the point of view of defensive purposes (much better places exist on the hills) and also from the point of view of maritime purposes (a much better and protected port exist on the other – south – side of the island, and from there one has an unobstructed view to Sicily and threats coming from this side). The place is unsuitable for an healthy living as it is too exposed to wind and sea-storms.

Hence, we propose that it may have been selected because it allowed a quite privileged position (Fig. 8) enabling to follow the rising and setting of the sun in the months close to the winter solstice, and – due to particular disposition of the two peaks - a relatively good determination of the solstice days. This is far from saying that the islanders used the Ustica horizon as a complete calendar in stone, since in all other days the sun rises and sets on the sea, but information about the arrival of midwinter and of the season bridging it might have been very useful for practical, especially navigational, purposes, as well as for religious ones. A confirmation of our – admittedly speculative – hypothesis may come from a currently ongoing extension of this study to contemporary sites of the same culture.

**Table 1.**

| Direction measured | Azimuth | Horizon height | Declination | Possible target | Difference |
|---|---|---|---|---|---|
| Main urban street | 162° | 3° | -44°30' | Rigil rises | 0° 19' |
| Center Village to summit of Monte Falconiera | 124° | 4° | -23°12' | Winter sosltice sunrise | 0° 38' |
| Center Village to easternmost point of Monte Falconiera | 114° | 0° | -19° | Sunrise Nov 18/ Jan 24 | - |
| Center Village to summit of Monte Costa del Fallo | 232° | 8° | -22° 58' | Winter sosltice sunset | 0° 52' |
| Center Village to westernmost point of Monte Costa del Fallo | 250° | 0° | -16° | Sunset Nov 6/ Feb 5 | - |

## ACKNOWLEDGEMENTS

The authors acknowledge the Authority to the Archaeological sites of Palermo, Sicily (Soprintendenza) for the kind permission to access to the site and in particular the Honorary Supervisor Vito Ailara for his kind help and support; as well as Dr. Cristiana Bonan and Mrs. Petra Licciardi for their technical support during photographic sessions.